# Crowdsourcing, Sharing Economies and Development

by

Araz Taeihagh

# Crowdsourcing, Sharing Economies and Development

Araz Taeihagh, School of Social Sciences, Singapore Management University, 90 Stamford Road
Level 4, Singapore, 178903 Phone: +65 68280627,
Email: araz.taeihagh@new.oxon.org

**ABSTRACT** – What are the similarities and differences between crowdsourcing and sharing economy? What factors influence their use in developing countries? In light of recent developments in the use of IT-mediated technologies, such as crowdsourcing and the sharing economy, this manuscript examines their similarities and differences, and the challenges regarding their effective use in developing countries. We first examine each individually and highlight different forms of each IT-mediated technology. Given that crowdsourcing and sharing economy share aspects such as the use of IT, a reliance on crowds, monetary exchange, and the use of reputation systems, we systematically compare the similarities and differences of different types of crowdsourcing with the sharing economy, thus addressing a gap in the current literature. Using this knowledge, we examine the different challenges faced by developing countries when using crowdsourcing and the sharing economy, and highlight the differences in the applicability of these IT-mediated technologies when faced with specific development issues.

## 1 Introduction

Crowdsourcing is the IT-mediated engagement of crowds for the purposes of problem-solving, task completion, idea generation and production (Howe 2006; 2008; Brabham 2008). Crowdsourcing encompasses various types of platforms, such as virtual labour markets (VLMs), tournament crowdsourcing (TC) and open collaboration (OC), which each have different roles and characteristics (Estellés-Arolas and González-Ladrón-de-Guevara 2012; Prpić, Taeihagh and Melton 2015). Along with the growth of crowdsourcing, another IT-mediated technology in the form of the sharing economy is rapidly being developed. 'Sharing economy' is an umbrella term referring to the practices of sharing, exchange or rental of goods and services to others through IT without the transfer of ownership. The sharing economy promises to increase efficiency and effectiveness by reducing transaction costs and increasing the rate of utilisation of goods and services. It has had a transformative effect on how goods and services are provided (Welsum 2016; Schor 2014; Goudin 2016).

Both crowdsourcing and the sharing economy are becoming increasingly popular (Lehdonvirta and Bright 2015; Cohen and Kietzmann 2014; Zervas, Proserpio and Byers 2016), but despite their rapid adoption and development there are gaps in the literature. These IT-mediated technologies improve efficiency and decrease transaction costs and information asymmetry, and share similarities in their use of IT, reliance on crowds, monetary exchange, use of reputation systems, etc. However, the literature in each domain tends to ignore the other or treats it as a singular form. Moreover, at times a platform is categorised as both a sharing economy platform and a crowdsourcing platform by different scholars. For instance, scholars distinguish between Amazon MTurk and TaskRabbit based on whether the task can be performed as a virtual service that can be executed online or whether a physical service needs to be performed locally (De Groen, Maselli and Fabo 2016; Aloisi 2015). However, there are many instances where both of these platforms have been categorised as part of the sharing economy. This issue is particularly prevalent when the topic under study relates to labour markets or commons.

In this work we aim to systematically compare the similarities and differences of various types of crowdsourcing and sharing economies across a wide range of criteria to address the gap in the literature and bring about a more nuanced understanding of these IT-mediated technologies. Furthermore, it is being suggested that developing countries can take up crowdsourcing and sharing economy platforms to address problems that are particular to development. In practice this can be difficult to achieve because developing countries face unique and specialized problems, and our knowledge about the various models of IT-mediated technologies is incipient. Thus, decisions are made by the market place despite or beyond the influence of policy makers.

The benefit of a more nuanced understanding of these IT-mediated technologies for developing countries is that industry and policy-makers can work together more effectively to leverage the new potential of applying IT-mediated technologies such as crowdsourcing and sharing economy for achieving development objectives while ensuring that they are implemented in ways that maximize positive impacts and minimize negative side effects. This paper thus suggests which forms would be more appropriate to which types of development issues, with particular focus on issues relating to mobile and online activities, productivity and innovation as well as legal and governance challenges.

In the next section we provide a general overview before examining different types of sharing economy and crowdsourcing in Section 3. Given that the sharing economy and crowdsourcing share characteristics we then systematically compare the two in Section 4. Then, drawing on this understanding, we examine different types of developing countries and focus on the challenges they face in regard to crowdsourcing and the sharing economy in Section 5, followed by concluding remarks in Section 6.

## 2 Background

### The sharing economy

The sharing economy is described as a transformative and disruptive economic model in which the consumption of physical goods, assets or services is carried out through rental, sharing or exchange of resources using IT through crowd-based services or intermediates without any permanent transfer of ownership (Lessig 2008; Botsman and Rogers 2010; Belk, 2014; Hamari et al. 2015; Dillahunt and Malone, 2015; Goudin 2016). This is done to increase efficiency and effectiveness by reducing transaction costs[1] and information asymmetry, particularly for consumers, increase the rate of utilisation of goods, recirculation of goods, exchange of services and sharing of productive assets, as well as increase competition in the marketplace, reduce the complacency of suppliers and make services that often exist in an informal fashion safer through formalisation (Goudin 2016; Schor, 2014; Welsum 2016; Hira and Reilly 2017).

Scholars have repeatedly criticised the term 'sharing economy' because it implies altruistic or positive non-reciprocal social behaviour that can increase societal trust and increased cooperation between individuals, when in fact the services involved are often fee-paying in nature and involve access to goods or assets that individuals often use for economic benefit (Belk 2014; Eckhardt and Bardhi 2015; Hamari et al. 2015).

---

[1] Demary (2015) based on Dahlman (1979) elaborates that platforms enable transaction costs to be reduced by facilitating: a) the finding of information and reduction of search costs; b) the checking of prices and decision making as well as bargaining on price; and c) a reduction in the policing and enforcement costs by enabling payments via the platform.

Based on the aforementioned description, the key features of the sharing economy are:[2]

- A transformative and disruptive nature, as evidenced by the effects of services such as Uber and Airbnb on the transportation and tourism sectors (Guttentag 2015; Ikkala and Lampinen 2015; Cannon and Summers 2014).
- The consumption and use of goods, services or assets through rental, sharing or exchange of resources, which increases the utilisation rate (Goudin 2016).
- A heavy reliance on IT through online platforms and/or mobile devices. For instance, the sharing economy relies on IT for identifying relevant individuals or businesses, exchanging and aggregating relevant information (e.g. products, services, usage), booking of services and payment of fees. Technological breakthroughs that have enabled such activities have reduced transaction costs and increased the reach of the sharing economy (Gansky 2010; Belk 2014; Goudin 2016).
- The direct engagement of crowds and/or intermediaries. The sharing economy focuses on consumer markets through crowd-based online services or intermediaries (Hamari et al. 2015) providing consumer-to-consumer (peer-to-peer) or business-to-consumer models. This particular aspect of the sharing economy in which economic activity is carried out through crowd-engagement directly connects to crowdsourcing (see Section 4). Moreover, a large portion of the communications happens through word of mouth and social media (Gansky, 2010).
- The temporary nature of the engagement (e.g. temporary transfer of ownership) (Belk 2014), rather than any permanent transfer of ownership of goods, distinguishes the sharing economy from e-commerce which is buying and selling of goods and services online (Burt and Sparks 2003).

## Crowdsourcing

Crowdsourcing is the IT-mediated engagement of crowds for the purposes of problem-solving, task completion, idea generation and production in which the dispersed knowledge of individuals and groups is leveraged through a mix of bottom-up innovative crowd-derived processes and inputs with efficient top-down goals set and initiated by an organisation (Howe 2006; 2008; Brabham 2008). It is continuously evolving and a variety of forms are emerging (for an in-depth review, see Prpić 2016 and Prpić & Shukla, 2016). Crowdsourcing initiatives can be carried out by 'propriety crowds' that organisations foster through their own in-house platforms or by using third-party crowdsourcing platforms that provide the required IT infrastructure and 'built-in crowds' as a paid service (Bayus 2013).

In this work, we use the three generalised types of crowdsourcing from the literature that focus on micro-tasking in Virtual Labour Markets, Tournament Crowdsourcing and Open Collaboration through social media and the web (Estellés-Arolas and González-Ladrón-de-Guevara 2012; Prpić et al. 2015). These general categorisations are not exclusive or exhaustive but they are useful for examining the general characteristics of different types of crowdsourcing and sharing economy.

## 3 Types of Sharing Economy and Crowdsourcing

Both sharing economy and crowdsourcing are umbrella terms and encompass a wide range of IT-mediated technologies which can be classified into different categories based on a diverse set of features and applications. Below we examine various types of sharing economy and crowdsourcing.

## Types of sharing economy

---

[2] As Gansky (2010) points out, not all of the characteristics listed above need to be present in every sharing economy business.

The sharing economy has the potential to be applied in a diverse range of sectors, which include[3]:

- Tourism and hospitality (Guttentag 2015; Ikkala and Lampinen 2015);
- Mobility and logistics (e.g. car-sharing, ride-sharing, bike-sharing and on-demand logistics and delivery) (Cohen and Kietzmann 2014; Cannon and Summers 2014; Techcrunch 2015);
- Labour and service platforms (Thompson 2015; Fraiberger and Sundararajan 2015);
- Food and dining (Hendrickson 2013; Tanz, 2014);
- Goods and equipment (Morrissey 2015; Long 2013; Anderson 2016);
- Financial services (Ordanini et al. 2011; Zhang et al. 2014); and
- Other rapidly developing new areas of application.

This sector-based categorisation is perhaps the easiest method of classification but, as Kenny, Rouvinen and Zysman (2015) point out, sectors are now blurring due to digitisation and use of IT platforms.

Belk (2010) suggests the concepts of 'sharing in' and 'sharing out' as a means of distinguishing between sharing that is similar to family sharing (ownership as common) and sharing with strangers that does not create any attachment or bonds. Demary (2015) reports on Smolka and Hienerth (2014) and their categorisation of a sharing economy based on whether transactions are market-mediated or not, while Kostakis and Bauwens (2014) and Oskam and Boswijk (2016) distinguish between types of sharing economy by focusing on whether the sharing economy platform is centrally controlled or open/decentralised and also whether the initiative is for profit or not. Cheng (2014) further expands the consideration of whether the platform is for-profit or not-for-profit and its distributed or centralised production aspects by also considering whether the application covers offline realms or not. Cheng (2014) makes a distinction between peer-to-peer platforms and other types of sharing economy business models but, as Westerbeek (2016) identifies, an overlap is still present between peer-to-peer platforms and these other types of business model (such as collaborative consumption and the gift economy).[4]

Botsman and Rogers (2010) take a functional approach and distinguish between three types of sharing economy based on whether the business model is: a) a redistribution market of used or pre-owned goods; b) a product service system where consumers pay for access to the good as a service rather than purchasing the good; and c) based on collaborative lifestyles (i.e. involving the sharing of non-physical assets such as time, expertise and space). Andersson et al. (2013) take a similar functional approach distinguishing business models based on peer-to-peer trading of digital and tangible materials, sharing of goods and sharing of services. They also examine the characteristics of sharing platforms based on the planning horizon for every transaction, assessing whether it is immediate (i.e. only a short time is required for planning every transaction), recurring (a long time is required for setting up the first transaction) or deferred (a long time is required for planning every transaction). Similarly, Demary (2015) characterises peer-to-peer platforms based on the cost of transactions (i.e. whether supplier, consumer, both supplier and consumer, or advertisers (in multisided platforms) pay the charges), while Choudary (2015) focuses on the architectural framework (using a categorisation based on whether the platform focuses on community building, the provision of infrastructure, or data) and patterns of exchange in the platforms (based on whether platforms primarily exchange information, currency, and/or goods and services).

In this study we adopt the sharing economy categorisation of Gansky (2010) and Rauch and Schleicher (2015). These authors consider two business models in which a business either owns goods or services and rents them out or creates a platform for the exchange of goods and services on a temporary basis and makes a profit by charging fees to parties involved in a myriad of ways (as illustrated earlier by Demary 2015). Gansky (2010) names these two sharing economy models the 'full mesh mode' (company assets rented out to customers) and the 'own-to-mesh mode' (platforms enabling peer-to-peer sharing of goods for a transaction or partnership fee rather than owning the goods). Rauch and Schleicher (2015) name these two business models 'asset hubs' (the business owns the goods or services and rents them out) or 'peer-to-peer sharing networks' (the business creates a peer-to-peer platform for the exchange of goods and services on a temporary basis).

---

[3] For a recent survey of adoption of sharing economy in developing countries see Hira (2017).
[4] Westerbeek (2016) defines peer-to-peer sharing as when the main objective of the business transaction can be reached using a one-on-one one-off transaction between a provider and a user (e.g. in an Uber ride a certain location is reached after a one-to-one transaction (the Uber ride)).

## Types of Crowdsourcing

### Virtual Labour Markets (VLMs)

A VLM is an IT-mediated market where individuals can provide online services that can be performed anywhere (often by engaging in spot labour), offered by organisations generally through micro-tasks, typifying the production model of crowdsourcing (Brabham 2008), in exchange for monetary compensation (Prpić, Taeihagh and Melton 2014; Luz, Silva and Novais 2015).

Micro-tasks offered at sites such as Amazon's Mechanical Turk (MTurk) and Crowdflower include document translation, transcription, photo and video tagging, editing, sentiment analysis, categorisation, data entry, and content moderation (Crowdflower 2016). These are activities that can be divided into various steps (micro-tasks) that can be completed in parallel and at scale using human computational power. Currently these tasks can be better performed through collective intelligence rather than through artificial intelligence and automation. Furthermore, at the moment most of the labourers working through VLM websites often work independently and anonymously and cannot form teams or groups using the VLM platforms. This is a function of the current design of these platforms and could (and probably will) change in the future to allow more sophisticated tasks to be performed. At the moment, most of these micro-tasks require low to medium levels of skill and are at times repetitive, meaning the compensation level per task is low.

### Tournament Crowdsourcing (TC)

TC (Zhang et al. 2015; Glaeser et al. 2016) is another form of crowdsourcing in which organisations post their problems to specialised IT-mediated platforms such as Eyeka or Kaggle or to in-house platforms such as Challenge.gov (Brabham 2013). Here, with the help of the IT-mediated platform, organisers form a competition and set the rules and prize(s) for the competition. Individuals or groups can post their solutions through the specialised IT-mediated platform to be considered for a prize, which range from a few hundred dollars to hundreds of thousands of dollars or even more.[5]

These TC platforms generally attract and maintain more specialised crowds who are interested in the particular focus of the platform, which can differ widely from computer science (Lakhani et al. 2010) and data science (Taieb and Hyndman 2014) to open government and innovation (The White House 2010). Relative to VLMs these TC platforms generally attract smaller numbers of more specialised individuals, and members can choose to not be anonymous at these sites so as to benefit from the reputational gains from their successful participations (Prpić, Taeihagh and Melton, 2015).

### Open Collaboration (OC)

In the OC model of crowdsourcing, problems or opportunities are posted by an organisation to the public through IT systems and crowds voluntarily engage in these endeavours generally without expecting monetary compensation (Michel, Gil and Hauder, 2015; Prpić et al. 2015). Examples of this type of crowdsourcing include starting an enterprise wiki or using social media and online communities to gain contributions (Crowley et al 2014; Budhathoki and Haythornthwaite 2013; Mergel, 2015).

---

[5] https://www.kaggle.com/competitions

The level of engagement from the crowd depends on a number of factors such the efficacy of the 'open call' by the organisation, the crowd capital of the organisation as well as the reach and engagement of the IT platform used (Prpić and Shukla 2013). As an example, as at May 2015, Twitter had more than 500 million users, out of which more than 310 million are active on a monthly basis.[6] However, this does not necessarily translate into significant engagement from the potential pool of users on the platform. An open call might get the attention of celebrities or Nobel Laureates and get significant traction and diffusion though their networks or, on the other hand, it might simply be ignored if the organisation does not have ample influence within the network. Factors such as level of popularity and level of prior engagement on the platform by the organisation, number of followers, number of retweets or mentions the organisation garners and popularity of the followers and their level of reach in turn, along with the content posted, are a small subset of the many factors that influence the level with which crowds might engage with the open call (Cha et al. 2010).

## 4 Comparison of crowdsourcing with the sharing economy

As mentioned in the introduction of the article, although various forms of the sharing economy and crowdsourcing can share a large number of common characteristics, the literature in each domain at times ignores the other or treats it as a singular form. Additionally, in some instances a platform is categorised as both a sharing economy platform and a crowdsourcing platform by different scholars. For instance, scholars generally distinguish between Amazon MTurk and TaskRabbit given that the former provides a virtual service that can be performed online and the latter provides a physical service that needs to be performed locally (De Groen, Maselli and Fabo 2016; Aloisi 2015). This distinction is fundamental to the works of Gansky (2010) and Rauch and Schleicher (2015) as they solely focus on the exchange of physical goods or services that must be provided in person, which implicitly differentiates between the sharing economy and crowdsourcing as crowdsourcing can be performed virtually. However, there are numerous instances in the literature where both of these platforms have been categorised as part of the sharing economy. This issue is particularly prevalent when the topic under study relates to labour markets or commons (e.g. Amazon MTurk and Wikipedia). Nevertheless, Westerbeek (2016) explicitly differentiates between crowdsourcing and sharing economy platforms by pointing out the one-on-one peer-to-peer aspect to be the most important part of the sharing economy that is *not* present in crowdsourcing.

As shown in sections 2 and 3 of this paper, crowdsourcing and the sharing economy both encompass a wide range of activities and business models. Crowdsourcing refers to three generalised types of VLM, TC and OC with varying levels of accessibility, crowd magnitude and scale as well as IT structure used (Prpić, Taeihagh and Melton 2015). Below we expand and enhance this characterisation of crowdsourcing types to cover the sharing economy in the form of asset hubs and peer-to-peer sharing networks and further consider platform architecture and interactions (see Table 1). By carrying out this systematic examination we address a key gap in the literature and bring to light a more nuanced picture of the similarities and differences between the crowdsourcing and sharing economy types. A quick examination of Table 1 shows that each of the five types of IT-mediated platforms have their own unique set of characteristics while sharing commonalities with the other four types of crowdsourcing and sharing economy platforms.

Accessibility

IT-mediated crowds can be examined based on the level of openness of their platform. Prpić et al. (2015) distinguish between platforms based on whether the platform is open to the public free of charge or requires payment for gaining access (and thus is private). OC platforms and peer-to-peer sharing networks are considered public, while TC, VLMs, and asset hubs are considered private (see Table 1).

Accessing peer-to-peer sharing networks such as Uber can be as simple as downloading an app on a mobile phone and a quick sign up, and in OC crowdsourcing similarly the payment of fees is not

---

required for accessing the service. Of course, the actual use of the service is an entirely different matter and often requires payments in peer-to-peer sharing networks – unlike OC crowdsourcing.[7]

In the case of TC and VLMs, individuals or organisations need to pay a launch fee to start a competition or access the spot labour (Prpić et al 2015). In a similar fashion, most for-profit asset hubs require the payment of a fee to access the service offered.[8]

**Table 1 - Comparison of different types of sharing economy and crowdsourcing**

|  | Accessibility | Anonymity | Crowd magnitude | Nature of the crowd | Platform architecture | IT-structure | Platform interactions |
|---|---|---|---|---|---|---|---|
| **Virtual labour markets** (e.g. Amazon Mturk) | Private | High | Millions | General | Community building and infrastructure provision | Episodic | Information, currency, and virtual services |
| **Tournament crowdsourcing** (e.g. Kaggle) | Private | Medium | Hundreds of thousands | Specialized | Community building | Episodic | Information, currency, and virtual services |
| **Open collaboration** (e.g. Twitter) | Public | Variable | Hundreds of millions | General | Community building | Collaborative | Information |
| **Asset hubs** (e.g. Zipcar, Car2go) | Private | Low | Hundreds of thousands to a few millions | Specialized | Community building and infrastructure provision | Episodic | Information and currency |
| **Peer-to-peer sharing networks** (e.g. Uber) | Mostly Public | Low | Hundreds of thousands to millions | General or specialized | Community building, infrastructure provision and data layer | Collaborative | Information and currency, in some instances goods/services as well |

Anonymity and reputation systems

Anonymity in the context of crowdsourcing and the sharing economy refers to whether the participants in the crowds in different types of platforms are anonymous with respect to their offline identity. OC platforms have a variable level of anonymity because of the different contexts and natures of the activities of a particular site, as well as user preferences (Prpić, Taeihagh and Melton 2015). VLMs such as Amazon MTurk provide 'methodological anonymity' by providing unique numeric identifiers to the requester as a means of connecting them with MTurk workers, which provides them with a high level of anonymity. TC platforms do not necessarily require the matching of offline and online identities, although strong incentives might exist for the crowds frequenting such sites to connect their offline and online identities to advance their offline career. Moreover, both crowdsourcing and sharing economy platforms use reputation systems to maintain and improve the participation of IT-mediated crowds. Morschheuser, Hamari and Koivisto (2016) review the use of reputation systems in crowdsourcing by examining the literature on the use of points/scores, leaderboards/rankings, badges/achievements, levels, progression and reward systems, etc. Furthemore, new studies on the use of reputation systems in the sharing economy are emerging (Slee 2013; Zervas et al. 2015; Ert et al. 2015).

---

[7] A variety of payment systems are used for transactions in peer-to-peer networks that range from payments from suppliers, consumers or both to payment by advertisers in multisided platforms (Demary 2015).

[8] For instance, in the case of car-sharing companies Car2Go has a $35 registration fee (plus tax) and Zipcar has a $25 one-time application fee (Car2Go 2016; Zipcar 2016). Both of these services offer plans catering to the needs of their members ranging from pay-as-you-go plans to monthly plans that offer certain prepaid miles that a member can use.

Crowd magnitude

Crowd magnitude refers to the number of available individuals to implement crowdsourcing or sharing economies by conducting activities such as performing a task or providing a service, which ultimately dictates the rate and scale with which resources can be created or provided in each platform (Prpić, Taeihagh and Melton 2015). Table 1 presents the magnitude of different crowds for each form of sharing economy and crowdsourcing reviewed.

Codagnone, Abadie and Biagi (2016) provide a review of the numbers of registered contractors on various sharing economy platforms, demonstrating that the size of largest crowds in peer-to-peer sharing networks can reach into the millions. In the case of asset hubs, Car2go has over a million users and Zipcar has close to a million users (Dryden 2015; Avis 2016). Prpić, Taeihagh, and Melton (2015) report on the largest size of crowds in crowdsourcing platforms, which range from thousands of participants to the hundreds of millions: OC platforms such as Twitter and Facebook have hundreds of millions of members, while TC platforms and VLMs' magnitude of crowds also range from hundreds of thousands (Kaggle, eYeka) to millions (Crowdflower, Amazon MTurk).

Nature of the crowd

In crowdsourcing and the sharing economy, specialised crowds form around specific types of content or service, while general crowds provide or perform a multitude of common tasks or services. The nature of the crowds can influence the size of the potential crowd available for a specific endeavour, as well as impacting the tasks assigned to the participants and the features of the IT used, for instance the various forms of TC, are unlikely to reach the same size as general OC platforms or some of the larger peer-to-peer sharing networks (Prpić et al. 2015).

Table 1 highlights that asset hubs and TC rely on specialised crowds, whereas OC crowdsourcing and VLMs rely on general crowds that either form around multiple kinds of content (OCs) or services (VLMs). Peer-to-peer sharing networks represent a more complex picture as their crowds can be specialised or general. For example, an individual or organisation interested in an asset hub such as Zipcar is largely interested in a specific type of good or service offered whereas in peer-to-peer sharing networks individuals might be interested in specific services (such as in the case of Uber) or be more generalised (such as in the case of TaskRabbit).

Platform architecture

Choudary (2015) examined a selection of IT platforms and categorised them based on their architectural frameworks and configuration and their patterns of exchange. He identified that all platforms function across three layers but the degree to which each layer is dominant varies:

1- Network-Marketplace-Community layer: comprises the individual members of the crowd and their network of interactions with other members. The network interaction might be direct with each other as in social networks or implicit in the case of markets in which buyers and sellers interact regularly. In some instance, this implicit community is formed when there are no direct interactions between the individual users but the platform leverages the data available from individual users and benchmarks them with one another to create value.

2- Infrastructure layer: enables value creation in the platform via the provision of tools, services and rules. The infrastructure system in itself does not create value but allows users to create value using this infrastructure, such as in the case of platforms such as YouTube that facilitate content creation, dissemination and monetisation.

3- Data layer: all platforms use data but the extent and intensity varies among them significantly. At a minimum, data is used for connecting the users of a platform with relevant goods/services/content. However, in some platforms data plays the leading role.

Table 1 highlights that OC, VLM and TC platforms all focus on value creation by creating a Network-Marketplace-Community layer. VLMs (more so than TC platforms) also focus on providing tools and services that facilitate the connection of individuals and organisations that demand work with crowds, as well as providing templates, tools and APIs that facilitate the creating of tasks and the receiving of results from the crowds. Asset hubs also strongly focus on the Network-Marketplace-Community layer as well as infrastructure provision. Companies such as Zipcar and Car2go, for instance, operate based on developing and maintaining sizeable fee-paying crowds and providing and maintaining an infrastructure network of vehicles for their use. Peer-to-peer sharing networks arguably have the most sophisticated architecture and rely on a mixture of Network-Marketplace-Community, infrastructure and data layers that vary in terms of the functionality they provide. Asset hubs also utilise data layers but given that asset hubs have more control over their own companies' assets relative to peer-to-peer sharing networks, which rely solely on users' goods/services, it can be argued that data layers are far more vital for the proper functioning of peer-to-peer networks. For instance, an asset hub such as Zipcar can relocate their own vehicles to different locations for the provision of service, whereas a company such as Uber has to utilise more sophisticate analytics to change the behaviour of their contracting drivers and provide coverage in different areas.

### Platforms' IT structure

Prpić and Shukla (2013) distinguish between two types of IT structure, namely collaborative IT structures and episodic IT structures, based on whether crowd members interact with each other through the IT platform for the purpose of deriving resources from the crowd. We can extend this concept to the sharing economy by examining whether IT-mediated crowds in the sharing economy need to interact with one another directly through the platform for the purpose of accessing goods or services (collaborative IT structures) or whether crowd members never need to directly interact with each other through the IT platform (episodic IT structures).

Prpić, Taeihagh and Melton (2015) highlight that VLMs use episodic IT structures (e.g. Amazon MTurk micro-tasks are carried out by individual crowd participants without interactions with each other, at least at the moment) and OC crowdsourcing platforms are found to generally use collaborative IT structures (e.g. social networks such as Twitter inherently exhibit collaborative IT structures due to extensive crowd interactions and over time), while TC platforms can allow both forms (e.g. an individual in a platform like kaggle can work separately from the others or can use the reputation system and results from previous competitions in the platform, connect with others and form teams for participating in the completion in the hope of increasing their chance of winning the tournament).

Similarly, in asset hubs, there is no need for the crowd members to connect with one another. For instance, crowd members using car-sharing services such as Zipcar or Car2go do not interact with one another and the central platform run by the asset holding company manages various coordination and scheduling efforts. Needless to say, the situation is completely different for peer-to-peer sharing networks as they directly rely on peer engagement and the collaborative aspect that the IT structure provides to function properly.

Table 1 illustrates that peer-to-peer sharing networks and OC crowdsourcing share similar collaborative IT structures while asset hubs along with VLMs and TC share similar episodic IT structures (not necessitating direct interaction of participants through the platform). As such, platforms that rely on collaborative IT structures require the existence, generation and maintenance of social capital to function properly (Prpić and Shukla 2013).

### Platform interactions

The dominant social and economic interactions in platforms revolve around the exchange of information, good/services or currency (Choudary 2015). All of the platforms highlighted in Table 1 share one fundamental aspect in that they all facilitate the exchange of information. VLMs, TC and asset hubs facilitate the exchange of information and currency in various forms. Furthermore, in VLMs and TC virtual services are also exchanged through the platform. Initially, the transfer of information from the individuals or organisations demanding work or expertise to workers and tournament participants is carried out. This is followed by the exchange of information and flow of virtual services in the form of the performance of tasks and provision of results and solutions.[9] Finally, a currency

---

[9] As described earlier, the distinction between virtual and physical services here is important. In VLMs and crowdsourcing, as the service can be performed online the flow of the service is virtual whereas in the sharing

exchange is carried out for the compensation of the crowd for their services. OC platforms are voluntary and often do not involve the exchange of currency or goods or services and thus the main form of exchange through such platforms is free information and/or content.[10]

Asset hubs and peer-to-peer sharing networks both involve the sharing of information and currency, generally through procedures such as: transfer of information on goods/services from provider (business or individual) to consumer, followed by the transfer of money from consumer to provider and subsequently the transfer of goods/services from the provider to the consumer. It is obvious that, unlike as is the case with virtual goods/services, in the case of physical assets the exchange of goods/services is not possible through the platform itself, although in some instances peer-to-peer sharing networks do also track, facilitate and monitor the exchange of goods/services internally. Choudary (2015) highlights that a Peer-to-peer sharing network such as Uber can track the 'transportation-as-a-service' exchange as it is aware of the path of the trip using GPS and mobile networks, which helps in terms of fee calculation and the determination of the completion of the ride.

## 5 Crowdsourcing, the sharing economy and development

The aim of the previous section was to bring attention to the nuanced similarities and differences between crowdsourcing and sharing economy platforms which can be used by developing countries when attempting to leverage these IT-mediated technologies for development. The comparison revealed that the five types of IT-mediated technologies examined (Asset Hubs and Peer-to-peer networks, Virtual labour markets, Tournaments crowdsourcing and Open collaboration) do not replicate each other and have unique attributes, while sharing commonalities with other forms. Understanding that developing countries have different development priorities helps in better capturing the challenges they face in the adoption of new technologies (Koch, 2015). One such approach that offers a more nuanced understanding of developing countries and their characteristics by taking into account countries' needs as well as resources and institutional capacities is the multi-dimensional clustering system of different types of developing countries. It categorises developing countries into five groups based on factors such as levels of poverty and inequality, productivity and innovation, political constraints, and dependence on external flows (Vázquez and Sumner, 2012; 2013; 2015). In this work each cluster of countries has a specific developmental character and set of issues that cannot be reduced to a simple representation using a single metric. In Table 2 we have developed a summary of the work by Vázquez and Sumner.

As with the introduction of any new technology, the proponents of crowdsourcing and the sharing economy focus on the positive aspects, such as the ease with which individuals can connect, interact, and exchange information, currency and goods and services, and promise positive societal transformation. While often the initial focus of scholars with the introduction of new technologies is on developed countries, developing countries can benefit from them as well. For instance, it is argued that sharing economy platforms, particularly peer-to-peer sharing networks, can boost small-scale service sectors in developing countries, as through the use of IT platforms they can reduce overhead costs and require relatively smaller levels of capital investment, solving informational problems by quickly matching consumers with suppliers (Ozimek 2014) and in fact recent survey data from Hira (2017) suggests an exponential increase in founding of sharing economy and crowdsourcing start-ups in developing countries.

While some scholars see the sharing economy and crowdsourcing as a potential pathway toward sustainability that can give voice to consumers and increase social capital, income and reciprocity,

---

economy platforms that entail the provision of physical services (e.g. TaskRabbit) the flow of service is not captured through the platform, meaning such platforms only allow the exchange of information and currency.

[10] It must be pointed out that solely the exchange of information is predominant in OC crowdsourcing platforms. In the sharing economy, it is possible that platforms solely facilitate the exchange of information, and exchange of goods/services and currency is carried out outside the platform (such as in the case of platforms that rely on advertisement and listing fees).

others warn of the potential for grave scenarios in which these platforms erode accountability and tax bases, divide communities, discriminate against individuals, underpay individuals, destroy job security, and result in the domination of markets by multinational corporations in the name of neoliberal capitalism (Heinrichs 2013; Dillahunt and Malone 2015; Martin 2016; Reeves 2015; Stone 2012; Edelman and Luca 2014; Hira and Reilly 2017). Similarly, Zvolska (2015) points out that while at the moment emphasis is generally placed on the potential sustainability of the sharing economy, concrete research substantiating these claims is scarce.

It must be pointed out that the success of development and the diffusion and use of innovative technologies depends on social, political and institutional factors (Edquist 2005; Schor, 2014). As was illustrated in the previous section, relative to their developed counterparts developing countries often fall behind in terms of GDP, levels of productivity, innovation, governance and political freedoms and have higher rates of poverty, income equality and dependence on external flows of cash. Given the nuanced differences within each country group, a one-size-fits-all approach to the adoption of crowdsourcing and the sharing economy in developing countries is not feasible. Below we focus on some of the relevant challenges facing different types of developing countries, with a particular focus on the governance and regulatory aspects.

**Table 2 Characteristics of different types of developing Countries – Developed based on Vázquez & Sumner research on groupings of developing countries (2012; 2013; 2015)[11]**

| | Poverty | Income inequality | Productivity | Innovation | GDP | Political freedom | Governance | CO2 emissions | External flow |
|---|---|---|---|---|---|---|---|---|---|
| **Type C1 - High poverty rate countries with largely traditional economies E.g. (2005-2010):** Sierra Leone; Ethiopia; Rwanda; Haiti; Bangladesh; Pakistan;India | Highest | Moderate | Lowest | Lowest | Lowest | Very Low | Poor | Low | High |
| **Type C2 – Natural resource dependent countries with little political freedom. E.g. (2005-2010):** Vietnam; Tajikistan; Yemen; Cameroon; Angola; Chad; Congo | High | Low | Low | Low | Low | Low - | Poor | Low | Moderate |
| **Type C3 - External flow dependent countries with high inequality E.g. (2005-2010):** Bolivia; Indonesia; Thailand; Peru; Colombia; Ukraine; Sri Lanka; Kenya | Moderate | High | Moderate | Moderate | Moderate | High | High | Moderate | High |
| **Type C4 - Economically egalitarian emerging economies with serious challenges of environmental sustainability and limited political freedoms E.g. (2005-2010):** Iraq; Egypt; China; Jordan; Azerbaijan; Venezuela | Moderate /Low | Lowest | High | High | High | Lowest | Poor | High | Low |
| **Type C5 - Unequal emerging economies with low dependence on external finance, E.g. (2005-2010):** Turkey; Brazil; Mexico; Argentina; South Africa; Malaysia | Lowest | High | Highest | Highest | Highest | Highest | Highest | Highest | Lowest |

---

[11] Vázquez & Sumner (2013) point out that even with this more nuanced categorization of the developing countries, it is not possible to perfectly match the group assignments of the countries. They point out that while Type C1 contains the most similar group of countries, the case of India is atypical (Gini coefficient considerably lower than the group average. GDP 16% higher in non-agricultural sectors relative to the group average. Lower exports of primary products, five times higher scientific article

Arguably the most important requirement for setting up and successfully operating crowdsourcing and sharing economy platforms in the first instance is access to communication networks for activities such as the exchange of information, currency and transactions among crowds (e.g. consumers with suppliers or workers with tasks from employers). According to Vázquez and Sumner's (2015) classification, which was elaborated in the previous section and in Table 2, Type C1 and C2 countries with the worst development indicators (i.e. higher levels of poverty and lower levels of labour productivity and innovation capacity) are dealing with severe poverty problems and have more difficulty in implementing such technologies to begin with. World Bank indicators on the diffusion of mobile phones by country groupings, mobile cellular subscriptions per 100 people, and individuals and households with access to internet suggest this is indeed the case (World Bank 2015; 2016).[12] Furthermore, Type C1 and C2 countries have higher levels of contribution from the agricultural sectors and larger portions of the population that have difficulty in using online platforms for carrying out more sophisticated tasks online such as participating in VLMs and TC that require higher capacity and access to computers rather than mobile phones that facilitate local (mobile) sharing economy activities (relative to their counterparts in C3 to C5 groupings, which have higher levels of urban population).

Research suggests that developed countries disproportionally hire more individuals from crowdsourcing and sharing economy platforms than developing countries to conduct online and local tasks (Codagnone, Abadie and Biagi 2016). Aside from issues relating to discrimination between individuals (discussed in the next subsection), here again the transfer of higher skilled and higher paying jobs within developing countries is not equal. Type C4 and C5 countries that generally have higher levels of productivity and innovation are more likely to get the better paying jobs such as programming and engage in specialised forms of IT-mediated technology such as TC. On the other hand, C1 to C3 countries will attract low to medium skilled work. Even in this case, Type C1 and to some extent C2 countries are at a disadvantage as it is more likely that individuals in these countries might not have the ability to provide verifiable personal information or demonstrate the lack of a criminal record (Nguyen, 2014) as part of joining a platform that might bar them from participating in online platforms as well as having more difficulty in transferring funds online. Therefore, it can be argued that, although a certain level of outsourcing from developed countries to developing countries is happening, the economies that have moved away from traditional agriculture and are more advanced will benefit more, which could in fact further increase the gap between C4 and C5 countries and their C1 to C3 counterparts that have more traditional economies.

<u>Governance</u>

Public governance is the process by which a society manages itself and organises its affairs and is a bedrock for successful and stable economies (UN, 2007). Developing countries often suffer from inefficiency in terms of the delivery of vital public services, inefficient revenue systems, poor transparency and the inappropriate allocation of resources, which often manifest themselves in acute problems in sectors such as healthcare (Shah, 2005; Berglof and Claessens 2006; Asante, Zwi and Ho 2006).

According to Ozimek (2014), poor governance and a lack of effective regulatory regimes in developing countries combined with weak property rights make attracting the investment required for building large companies with high reputational capital difficult. He argues that in the absence of good governance practices, nimble decentralised crowd-based rating systems lower the bar for the existence of an effective services industry and bypass the need for regulation, as users in these countries will trust peer-based feedback systems that can inform them about quality of goods/services more that

---

production and four times lower dependence on external finance relative to the group average as well as better governance and democracy indicators).

[12] The most important difference between C1 and C2 countries according to Vázquez and Sumner (2013) is in terms of level of primary exports (much higher in C2), quality of democracy (higher in C1) and dependency on external finance (higher in C1).

government endorsed companies and will help them in avoiding fraud and wrongdoing. However, Aloisi (2015) believes these ranking systems and approval ratings transfer the traditional role of management to the users of the platforms, highlighting that with this transfer the recipients of such reviews in the platforms are less protected from external manipulation and agendas. Furthermore, given that most of crowdsourcing and sharing economy companies are commercial and seek profits (with the exception of some OC crowdsourcing platforms and non-commercial peer-to-peer sharing networks) Ozimek's views about potential of IT platforms seem rather optimistic.

Codagnone, Abadie and Biagi (2016) have already documented instances of litigation in the US in regard to crowdsourcing and the sharing economy concerning employee benefits, cost reimbursements, violation of labour standards, incorrect classification as contractors, and minimum wage and overtime payments. Uncontrolled price wars between firms can also affect workers, employees and contractors. Stiff competition can result in price reductions by firms seeking to attract more consumers and an increasing volume of business but this can also result in contracting drivers being undermined, affecting the industry and ultimately consumers as a whole (Straits Times 2016). If such issues are surfacing so quickly with the adoption of crowdsourcing and sharing economy practices in developed countries such as the US and Singapore, which have strong governance and regulatory regimes, as well as effective enforcement mechanisms relative to developing countries, the counter argument that given the governance and regulatory deficits in developing countries a stronger and stricter enforcement and oversight of these platforms is needed also seems plausible.

In developed countries, in response to some of these legal challenges, firms such as TaskRabbit, Uber and Lyft have made adjustments to their business models. However, without adequate regulation being in place, Type C1, C2 and C4 countries are susceptible to firms entering their markets and dominating them while passing on the risks to workers, contractors and consumers (e.g. not having strict regulations for mandating third-party insurance in ride-sharing platforms or protecting privacy and financial information in both commercial crowdsourcing and sharing economy platforms that carry out currency exchanges) and then dealing with any litigation afterwards, perhaps after a long period in which they took advantage of the situation. This is further exacerbated because these countries (particularly C1 and C2 types) are less capable of monitoring the activities of the platforms and ensuring the correct collection of records and sufficient tax payments to the state.

Codagnone, Abadie and Biagi (2016) and Aloisi (2015) focus on the work-related challenges surrounding IT-mediated platforms, examine the relevant literature and meticulously unpack issues such as workplace health and safety, discrimination, and social arbitrage to address exploitation using these platforms and facilitate employment online (e.g. Amazon MTurk) or locally (e.g. TaskRabbit). Accordingly, they suggest:

- The development of a minimum wage and maximum working hours per day restrictions;
- Avoiding exclusivity clauses that tie workers to a particular platform;
- The inclusion of relevant forms of social protection and health insurance;
- The provision of liability insurance for damage to third parties;
- Privacy protection mechanisms for workers;
- Guarantees for avoiding algorithmic discrimination with respect to geographical preferences, gender, ethnicity, race or age when matching individuals in platforms; and
- Mandating the portability of an individual's performance across platforms.

Many of the suggested remedies are challenging to implement and are yet to be addressed in developed countries, which further increases concerns in regard to developing countries. All of the developing countries can benefit from improving their governance and regulatory capacity and capability relative to developed countries. This in turn will facilitate addressing the aforementioned issues. As highlighted in Table 2, Type C1, C2 and C4 countries have the highest levels of governance deficit, which demonstrates the challenges in addressing the issues raised by Codagnone, Abadie and Biagi (2016) and Aloisi (2015). Moreover, it is worth pointing out that the compounding effects of corruption and restrictions on political freedom in these countries further exacerbate these problems, as such workers in these countries will be more vulnerable relative to their Type C3 and C5 counterparts.

Highly publicised concerns about Uber, for instance, due to excessive charges from the surge-pricing algorithm and drivers being accused of assault, resulted in blanket bans in some cities, as unlike the traditional taxi industry Uber was initially not subject to strict regulations for pricing and licensing (Gobble, 2015). However, the findings of the study by van den Broek (2015) suggest that, although firms such as Airbnb and Uber try to hold on to their generic business models as much as possible, in the face of regulatory constrains (mainly relating to drivers in the case of Uber and hosts in the case of Airbnb) these firms have adapted their business models and have found ways to operate legally within the set framework.

As such, the active participation of governments in developing countries and more effective regulation of the affected sectors is paramount for gaining the benefits of IT-mediated platforms highlighted earlier (e.g. even addressing shortcomings in provision of goods/services by the state as suggested by Ozimek (2014)) and avoiding negative consequences such as labour law violations, discrimination, infringements on privacy, etc. Type C3 and C5 countries, with higher governance capacities, are more likely to be able to work with firms, or impose restrictions on them to encourage the adoption of positive practices. Additionally, given the higher level of productivity in Type C4 and C5 countries, they can utilise pull mechanisms to direct innovation in IT-mediated technology and provide funding and support to companies that follow best practices. Research by Sadoi (2008) suggests that focusing on developing local technological capabilities within a country is more successful than the provision of incentives to firms for technology transfer to developing countries as successful transitions depend on countries developing their own innovation hubs. As such, Type C1 to C3 countries should not just open markets to external corporations but should exert some control and focus on improving levels of productivity and innovation and perhaps given the complexity of the issues at hand and the severity of constrains they face set stricter control mechanisms relative to their C4 and C5 counterparts, or even focus on direct provision of services.

It is worth mentioning that some forms of crowdsourcing platform, particularly OC crowdsourcing, rather than receiving support, might be strictly limited in some of the developing countries with lower levels of political freedom (Type C4, C1 and C2) or actively used for reducing political freedom, as new empirical research by Asmolov (2015) demonstrates that, using volunteers from crowdsourcing platforms, it is possible to prevent collective action.

## 6 Conclusion

This paper examined the sharing economy and crowdsourcing and highlighted various types of each IT-mediated technology. Afterwards, given the similarities of crowdsourcing and sharing economy in terms of their use of IT, reliance on crowds, monetary exchange, use of reputation systems and the gap in the literature in regard to their nuanced similarities and differences, the sharing economy and crowdsourcing were systematically compared along several dimensions.

We advanced the literature on the sharing economy and crowdsourcing by providing a comparison of their types across dimensions such as accessibility, crowd magnitude, nature of the crowd, anonymity, platforms' architectural frameworks, IT structure and interactions. This systematic comparison brought a more nuanced understanding of these IT-mediated technologies and highlighted similarities and differences between various types of crowdsourcing and sharing economy platforms, demonstrating that each type of IT-mediated technology examined had unique attributes, while sharing commonalities with other forms. It addressed a gap in the current literature where these IT-mediated technologies were either ignored in the other domain or treated as a singular form or even at times categorised as both a sharing economy platform and a crowdsourcing platform by different scholars. Of course, examinations across these dimensions each include exceptions and the comparison should not be considered definitive. Nevertheless, it allows future researchers to better differentiate between the crowdsourcing and sharing economy. For instance, following Gansky (2010) and Rauch and Schleicher (2015) we mainly focused on the commercial use of the sharing economy that focuses on the exchange of physical goods or services that are carried out locally. This can be expanded in the future once research in the field goes beyond the types of categorisations highlighted in the current paper and coalesces around a more detailed categorisation of sharing economy types. This endeavour is currently under development, particularly in regard to peer-to-peer networks.

In addition to the above contributions, we examined the use of crowdsourcing and the sharing economy in developing countries. We went beyond the simple categorisation of developing economies based on GDP and examined some of the challenges facing different groups of developing countries in addressing crowdsourcing and the sharing economy. We suggested which forms would be more appropriate when faced with different types of development issues, with particular focus on issues relating to mobile and online activities, productivity and innovation as well as legal and governance challenges, helping to highlight the differences in the applicability of these IT-mediated technologies in specific contexts.

We hope that this research facilitates a more nuanced examination of the applicability of these technologies in different types of developing countries, encourages researchers to study them more rigorously in future, and helps industry and policy-makers to work together more effectively to leverage the new potential of applying crowdsourcing and sharing economy for addressing developmental challenges while maximizing positive impacts and minimizing negative side effects.

## References

Aloisi, A.(2015). Commoditized Workers: Case Study Research on Labour Law Issues Arising from a Set of 'On-Demand/Gig Economy' Platforms. http://ssrn.com/abstract=2637485

Anderson R.(2016). How some big firms are learning to share as well as sell. BBC http://www.bbc.com/news/business-35362199

Andersson, M., Hjalmarsson, A. and Avital, M.(2013). Peer-to-Peer Service Sharing Platforms: Driving Share and Share Alike on a Mass-Scale. 34th International Conference on Information Systems, Milan.

Asante, A.D., Zwi, A.B. and Ho, M.T.(2006). Equity in resource allocation for health: A comparative study of the Ashanti and Northern Regions of Ghana. *Health Policy*, 78(2):135–148.

Asmolov, G.(2015). Vertical Crowdsourcing in Russia: Balancing Governance of Crowds and State–Citizen Partnership in Emergency Situations. *Policy & Internet* 7(3):292–318.

Avis (2016). First Quarter 2016 Earnings Call. Avis Budget Group. http://ir.avisbudgetgroup.com/common/download/download.cfm?companyid=ABEA-36XVJR&fileid=889543&filekey=7F284FF5-94E7-4F77-B5A0-3161CCC07B16&filename=Avis_Budget_Earnings_Presentation_-_May_2016_-_FINAL.pdf

Bayus, B.L.(2013). Crowdsourcing new product ideas over time: An analysis of the Dell IdeaStorm community. Management science, 59(1):226-244.

Belk, R.(2010). Sharing. *Journal of Consumer Research*, 36(5):715–734.

Belk, R.(2014). You are what you can access: Sharing and collaborative consumption online. *Journal of Business Research*, 67(8):1595–1600.

Berglof, E. and Claessens, S.(2006). Enforcement and good corporate governance in developing countries and transition economies, *World Bank Research Observer*, 21(1):123–150.

Botsman, R. and Rogers, R.(2010). *What's Mine is Yours*. Collins.

Brabham, D.C.(2008). Crowdsourcing as a model for problem solving an introduction and cases. *Convergence*. 14(1):75–90.

Brabham, D.C.(2013). Using Crowdsourcing in Government. IBM Center for The Business of Government.

http://www.businessofgovernment.org/sites/default/files/Using%20Crowdsourcing%20In%20Government.pdf

Budhathoki, N.R. and Haythornthwaite, C.(2013). Motivation for open collaboration crowd and community models and the case of OpenStreetMap. *American Behavioral Scientist*, 57(5):548–575.

Burt, S. and Sparks, L. (2003). E-commerce and the retail process: a review. *Journal of Retailing and Consumer Services*, 10(5), 275–286.

Cannon, S. and Summers, L.H.(2014). How Uber and the Sharing Economy Can Win Over Regulators. *Harvard Business Review*, 13.

Car2go (2016). Signup Page. Retrieved June 2016 http://www.car2go.com

Cha, M., Haddadi, H., Benevenuto, F. and Gummadi, P.K.(2010). Measuring User Influence in Twitter: The Million Follower Fallacy. ICWSM, 10(10−17):30.

Cheng, D.(2014). Is sharing really caring? A nuanced introduction to the peer economy. Open Society Foundation Future of Work Inquiry.
http://static.opensocietyfoundations.org/misc/future-of-work/the-sharing-economy.pdf

Choudary, S.P.(2015). Platform Scale: How an Emerging Business Model Helps Startups Build Large Empires with Minimum Investment. Platform Thinking Labs

Codagnone, C., Abadie, F. and Biagi, F.(2016). The Future of Work in the 'Sharing Economy'. Market Efficiency and Equitable Opportunities or Unfair Precarisation? Institute for Prospective Technological Studies, http://ssrn.com/abstract=2784774

Cohen, B. and Kietzmann, J.(2014). Ride on! Mobility business models for the sharing economy. *Organization & Environment*, 27(3):279–296.

Crowdflower (2016). Crowdflower Inc. Retrieved May 2016 www.clowdflower.com

Dahlman, C.J.(1979). The Problem of Externality. *Journal of Law and Economics*, 22(1):141–162.

De Groen, W.P., Maselli, I. and Fabo, B.(2016). The Digital Market for Local Services: A one-night stand for workers? No.133. CEPS Special Report.

Demary, V.(2015). Competition in the sharing economy No.19. IW policy paper.

Dillahunt, T.R., Malone, A.R.(2015). The Promise of the Sharing Economy among Disadvantaged Communities. *ACM Human Factors in Computer Systems*.
http://www.tawannadillahunt.com/wp-content/uploads/2012/12/pn0389-dillahuntv2.pdf

Dryden (2015). Car2go car-sharing service to suspend South Bay operations. DailyBreeze.com http://www.dailybreeze.com/business/20150504/car2go-car-sharing-service-to-suspend-south-bay-operations

Eckhardt, G.M., Bardhi, F.(2015). The Sharing Economy Isn't About Sharing at All. *Harvard Business Review*. https://hbr.org/2015/01/the-sharing-economy-isnt-about-sharing-at-all

Edelman, B.G. and Luca, M.(2014). Digital discrimination: The case of airbnb. com. *Harvard Business School NOM Unit Working Paper* (14-054).

Edquist, C.(2005): Systems of Innovation. Perspectives and Challenges. In: Fagerberg, J., Mowery, D.C. and Nelson, R.R.(eds.). *The Oxford Handbook of Innovation*. Oxford University Press, 181–208.

Ert, E., Fleischer, A. and Magen, N.(2015). Trust and Reputation in the Sharing Economy: The Role of Personal Photos on Airbnb. http://ssrn.com/abstract=2624181

Estellés-Arolas, E. and Gonzalez-Ladron-De-Guevara, F.(2012). Towards an integrated crowdsourcing definition. *Journal of Information Science*. 38(2):189–200.

Fraiberger, S.P. and Sundararajan, A.(2015). Peer-to-peer rental markets in the sharing economy. *NYU Stern School of Business Research Paper*. Chicago

Gansky, L.(2010). *The Mesh: Why the future of business is sharing*. Penguin.

Glaeser, E.L., Hillis, A., Kominers, S.D. and Luca, M.(2016). Crowdsourcing City Government: Using Tournaments to Improve Inspection Accuracy. *The American Economic Review*, 106(5):114–118.

Gobble, M.M.(2015). Regulating innovation in the new economy. *Research-Technology Management*, 58(2):62–64.

Goudin, P.(2016). The cost of non-Europe in the sharing economy: Economic, social and legal challenges and opportunities. European Parliament, January.

Guttentag, D.(2015). Airbnb: disruptive innovation and the rise of an informal tourism accommodation sector. *Current Issues in Tourism*, 18(12):1192–1217.

Hamari, J., Sjöklint, M. and Ukkonen, A.(2015). The sharing economy: Why people participate in collaborative consumption. *Journal of the Association for Information Science and Technology*.

Heinrichs, H.(2013). Sharing economy: a potential new pathway to sustainability. *GAIA-Ecological Perspectives for Science and Society*, 22(4):228–231.

Hendrickson (2013). Lining Up to Dine With Strangers: EatWith Pairs Local Hosts With Diners From All Over the World, *Wall Street Journal*. http://www.wsj.com/news/articles/SB10001424127887324085304579007092945525818

Hira A.(2017). Profile of the Sharing Economy in the Developing World: Examples of Companies Trying to Change the World, *Journal of Developing Societies*.

Hira, A., Reilly K.(2017) The Emergence of the Sharing Economy: Implications for Development, *Journal of Developing Societies*.

Howe, J.(2006). Crowdsourcing: A Definition. Crowdsourcing.com. http://crowdsourcing.typepad.com/cs/2006/06/crowdsourcing_a.html

Howe, J.(2008). Crowdsourcing: Why the Power of the Crowd is Driving the Future of Business. Crown Business. http://www.bizbriefings.com/Samples/IntInst%20---%20Crowdsourcing.PDF

Ikkala, T. and Lampinen, A.(2015). Monetizing network hospitality: Hospitality and sociability in the context of AirBnB. 18th ACM Conference on Computer Supported Cooperative Work & Social Computing (1033–1044).

Kenney, M., Rouvinen, P. and Zysman, J.(2015). The Digital Disruption and its Societal Impacts. *Journal of Industry, Competition and Trade*, 15(1):1–4.

Koch, S.(2015). From poverty reduction to mutual interests? The debate on differentiation in EU development policy. *Development Policy Review*, 33(4):479–502.

Kostakis, V. and Bauwens, M.(2014). Network society and future scenarios for a collaborative economy. Palgrave Macmillan.

Lakhani, K., Garvin, D.A. and Lonstein, E.(2010). Topcoder: Developing software through crowdsourcing. *Harvard Business School Management Case* (610–032).

Lehdonvirta, V. and Bright, J.(2015). Crowdsourcing for public policy and government. *Policy & Internet*, 7(3):263–267.

Lessig L.(2008). Remix: making art and commerce thrive in the hybrid economy. Penguin.

Long C.(2013). The gifts that keep on giving, *Sydney Morning Herald*. http://www.smh.com.au/money/saving/the-gifts-that-keep-on-giving-20131130-2yii9.html

Luz, N., Silva, N. and Novais, P.(2015). A survey of task-oriented crowdsourcing. *Artificial Intelligence Review*, 44(2):187–213.

Martin, C.J.(2016). The sharing economy: A pathway to sustainability or a nightmarish form of neoliberal capitalism? *Ecological Economics*, 121:149–159.

Mergel, I.(2015). Open collaboration in the public sector: The case of social coding on GitHub. *Government Information Quarterly*, 32(4):464–472.

Michel, F., Gil, Y. and Hauder, M.(2015). A virtual crowdsourcing community for open collaboration in science processes. Americas Conference on Information Systems.

Morrissey J.(2015). Sharing Economy Goes Hyperlocal With a Growing Market for Household Items, *NY Times*. http://www.nytimes.com/2015/09/03/business/smallbusiness/sharing-economy-goes-hyperlocal-with-a-growing-market-for-household-items.html?_r=0

Morschheuser, B., Hamari, J. and Koivisto, J.(2016). Gamification in crowdsourcing: a review. 49th Hawaii International Conference on System Sciences, pp.4375–4384.

Nguyen, G.T.(2014). Exploring collaborative consumption business models-case peer-to-peer digital platforms, Master's Thesis, Aalto University.

Ordanini, A., Miceli, L., Pizzetti, M. and Parasuraman, A.(2011). Crowd-funding: transforming customers into investors through innovative service platforms. *Journal of Service Management*, 22(4):443–470.

Oskam, J. and Boswijk, A.(2016). Airbnb: the future of networked hospitality businesses. *Journal of Tourism Futures*, 2(1):22–42.

Ozimek, A.(2014). The Sharing Economy and Developing Countries. *Forbes Business*. http://www.forbes.com/sites/modeledbehavior/2014/08/04/the-sharing-economy-and-developing-countries/#7a936cac56a4

Prpić J., Melton J., Taeihagh A. and Anderson T.(2015). MOOCs and crowdsourcing: Massive courses and massive resources. *First Monday*, 20(12). http://dx.doi.org/10.5210/fm.v20i12.6143

Prpić, J. and Shukla, P.(2013). The Theory of Crowd Capital. 46th Hawaii International Conference on System Sciences, Computer Society Press.

Prpić, J., Taeihagh, A. and Melton, J.(2014). Experiments on Crowdsourcing. Policy Assessment. Proceedings of IPP 2014. Oxford Internet Institute.

Prpić, J., Taeihagh, A. and Melton, J.(2015). The Fundamentals of Policy Crowdsourcing. Policy & Internet, 7:340–361. doi:10.1002/poi3.102

Prpić, J.(2016). Next Generation Crowdsourcing for Collective Intelligence. *Collective Intelligence*.

Prpić, J., & Shukla, P.(2016). Crowd Science: Measurements, Models, and Methods. 49th Hawaii International Conference on System Sciences, pp.4365-4374.

Rauch, D.E. and Schleicher, D.(2015). Like Uber, But for Local Governmental Policy: The Future of Local Regulation of the "Sharing Economy". *George Mason Law & Economics Research Paper*, (15–01).

Reeves, R.(2015). The free worker forlorn hope? *Management Today*, 36.

Sadoi Y.(2008). Technology transfer in automotive parts firms in China. *Asia Pacific Business Review*. 14(1):147–163.

Schor, J.(2014). Debating the Sharing Economy. *Great Transition Initiative*. http://www.greattransition.org/images/GTI_publications/Schor_Debating_the_Sharing_Economy.pdf

Shah, A.(2005). Public services delivery. World Bank Publications.

Slee, T. (2013). Some Obvious Things About Internet Reputation Systems, Working Paper.

Small, T.A.(2012). eGovernment in the Age of Social Media: An Analysis of the Canadian Government's Use of Twitter. *Policy & Internet*, 4(3–4):91–111.

Smolka, C. Hienerth, C.(2014). The Best of Both Worlds: conceptualizing trade-offs between Openness and Closedness for Sharing Economy Models.

Stone, B.(2012). My life as a TaskRabbit. *Bloomberg Businessweek*. http://www.bloomberg.com/news/articles/2012-09-13/my-life-as-a-taskrabbit

Straits Times (2016). Fare cuts by Uber, Grab will hurt sector: Taxi body http://www.straitstimes.com/singapore/fare-cuts-by-uber-grab-will-hurt-sector-taxi-body

Taieb, S.B. and Hyndman, R.J.(2014). A gradient boosting approach to the Kaggle load forecasting competition. *International Journal of Forecasting*, 30(2):382–394.

Tanz, J.(2014). How Airbnb and Lyft Finally Got Americans to Trust Each Other. *Wired Magazine*. http://www.wired.com/2014/04/trust-in-the-share-economy/

Techcrunch (2015). On-Demand Logistics Startup Lalamove Lands $10M to Fuel its China Expansion. http://techcrunch.com/2015/09/09/on-demand-logistics-startup-lalamove-scores-10m-to-fuel-its-china-expansion

The White House (2010). Guidance on the Use of Challenges and Prizes to Promote Open Government. https://www.whitehouse.gov/sites/default/files/omb/assets/memoranda_2010/m10-11.pdf

Thompson, M. (2015). Sharing economy makes it pay to work on your own. *CNBC*. http://www.cnbc.com/2015/07/01/sharing-economy-makes-it-pay-to-work-on-your-own.html

UN (2007). Public Governance Indicators: A Literature Review. New York: United Nations.

van den Broek, D.(2015). An exploratory research on the adaptation of sharing economy business models in different institutions. Master's thesis, University of Amsterdam.

Vázquez, S.T. and Sumner, A.(2012). Beyond Low and Middle Income Countries, IDS Working Papers, (404):1–40.

Vázquez, S.T. and Sumner, A.(2015). Is the 'Developing World' Changing? *European Journal of Development Research*.

Vázquez, S.T. and Sumner, A.(2013). Revisiting the meaning of development. *Journal of Development Studies*, *49*(12):1728–1745.

Welsum, D.V.(2016). Sharing is Caring? Not quite. Some observations about the 'sharing economy'. *World Development Report 2016,* World Bank.

Westerbeek, J.B.(2016). Mapping the Effects of Peer-to-Peer Sharing Economy Platforms on Society, Doctoral dissertation, TU Delft.

World Bank (2015). *The little data book on information and communication technology*. Washington D.C
https://openknowledge.worldbank.org/bitstream/handle/10986/22554/9781464805585.pdf

World Bank (2016). *The little data book*. Washington D.C

Zervas, G., Proserpio, D. and Byers, J.(2015). A first look at online reputation on Airbnb, where every stay is above average.

Zervas, G., Proserpio, D. and Byers, J.(2016). The rise of the sharing economy: Estimating the impact of Airbnb on the hotel industry. Boston University School of Management Research Paper (2013–16).

Zhang, N., Datta, S. and Kannan, K.N.(2014). An Analysis of Incentive Structures in Collaborative Economy: An Application to Crowdfunding Platform.
http://ssrn.com/abstract=2518662

Zhang, Y., Gu, Y., Song, L., Pan, M., Dawy, Z. and Han, Z.(2015). Tournament Based Incentive Mechanism Designs for Mobile Crowdsourcing. IEEE Global Communications Conference (1–6).

Zipcar (2016). Rates and Plans. Zipcar.com. Retrieved June 2016.

Zvolska, L.(2015). Sustainability Potentials of the Sharing Economy: The case of accommodation sharing platforms. IIIEE Master's thesis.